\documentclass[unnumsec,webpdf,contemporary,large]{oup-authoring-template}

\usepackage[utf8]{inputenc}
\usepackage{acronym}
\usepackage{graphicx}
\usepackage{hyperref}

\begin{document}

\journaltitle{Microscopy and Microanalysis}

\copyrightyear{2023}

\appnotes{Paper}

\firstpage{1}

\title{Live Iterative Ptychography}

\author[1, $\ast$]{Dieter Weber\ORCID{0000-0001-6635-9567}}
\author[4,5]{Simeon Ehrig\ORCID{0000-0002-8218-3116}}
\author[2,3]{Andreas Schropp\ORCID{0000-0000-0000-0000}}
\author[1]{Alexander Clausen\ORCID{0000-0002-9555-7455}}
\author[2]{Silvio Achilles\ORCID{0000-0001-7244-7854}}
\author[4]{Nico Hoffmann}
\author[4]{Michael Bussmann\ORCID{0000-0002-8258-3881}}
\author[1]{Rafal E.~Dunin-Borkowski\ORCID{0000-0001-8082-0647}}
\author[2]{Christian G.~Schroer\ORCID{0000-0002-9759-1200}}

\address[1]{\orgdiv{Ernst Ruska-Centre for Microscopy and Spectroscopy with Electrons}, \orgname{Forschungszentrum J\"ulich}, \orgaddress{\postcode{52425 J\"ulich}, \country{Germany}}}
\address[2]{\orgdiv{Center for X-ray and Nano Science CXNS}, \orgname{Deutsches Elektronen-Synchrotron DESY}, \orgaddress{\street{Notkestr. 85}, \postcode{22607 Hamburg}, \country{Germany}}}
\address[3]{\orgdiv{Helmholtz Imaging}, \orgname{Deutsches Elektronen-Synchrotron DESY}, \orgaddress{\street{Notkestr. 85}, \postcode{22607 Hamburg}, \country{Germany}}}
\address[4]{\orgname{Helmholtz-Zentrum Dresden-Rossendorf}, \orgaddress{\postcode{D-01328 Dresden}, \country{Germany}}}
\address[5]{\orgname{Center for Advanced Systems Understanding}, \orgaddress{\street{Untermarkt 20}, \postcode{02826 G\"orlitz}, \country{Germany}}}

\corresp[$\ast$]{Corresponding author. \href{email:d.weber@fz-juelich.de}{d.weber@fz-juelich.de}}

\abstract{
    We demonstrate live-updating ptychographic reconstruction with ePIE, an iterative ptychography method, during ongoing data acquisition. The reconstruction starts with a small subset of the total data, and as the acquisition proceeds the data used for reconstruction is extended. This creates a live-updating view of object and illumination that allows monitoring the ongoing experiment and adjusting parameters with quick turn-around. This is particularly advantageous for long-running acquisitions. We show that such a gradual reconstruction yields interpretable results already with a small subset of the data. We show simulated live processing with various scan patterns, parallelized reconstruction, and real-world live processing at the hard X-ray ptychographic
    nanoanalytical microscope PtyNAMi at the PETRA III beamline.
}


\maketitle

\section{Introduction}

\begin{figure}
    \centering
    \includegraphics[width=230pt]{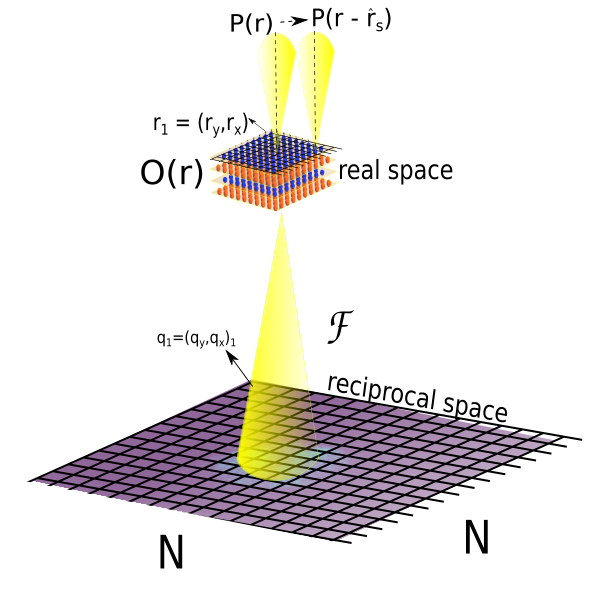}
    \caption{Schematic setup for far field ptychography. The illumination is localized and scanned over the object O(r). The detector captures the transmitted beam in the far field for each scan position. Illustration based on~\citep{BangunWDD}.}
    \label{fig:ptycho}
\end{figure}

Ptychography~\citep{Hoppe1969, Hoppe1969a, Hoppe1969b} can reconstruct a quantitative object transfer function for a specimen from a set of measurements where a spatially modulated illumination is shifted relative to the object and the intensity distribution of the transmitted and projected beam, typically far-field diffraction patterns, are measured for the different shift positions (Figure~\ref{fig:ptycho}). The alternating projection method, i.e., PIE/ePIE reconstruction algorithm~\citep{rodenburg2004phase,MAIDEN20091256} is one of the classical and well-known approaches to perform this reconstruction.

This method is currently in use at the hard X-ray ptychographic nanoanalytical microscope PtyNAMi at the PETRA III beamline P06 at DESY~\citep{Schropp:dy5005}. It allows to retrieve images of objects at a higher spatial resolution than the size of the X-ray focus and is physically only limited by the collection angle, wave length of the probe, and the available coherent photon flux~\citep{Schropp_2010}.

Here, a ptychographic data set is used for live processing that is recorded in the standard data format available the beamline. The raw data including the diffractions patterns and position values are stored in individual hdf5 files following in general the NeXus format.

Currently, the data is exported from the the standard data format available at the beamline, and then processed offline using a stand-alone software. The raw data, including the diffractions patterns and position values, are stored in individual HDF5 files following, in general, the NeXus format~\citep{Koennecke2015}.~\citep{Weber2023} contains the dataset ``scan\_00063'' with such a raw beamline data structure. Adjusting the parameters for ptychography requires a range of test measurements to adjust exposure, beam position on the detector, specimen region, focal distance and settings for the ptychography engine. Since many specimens exhibit mostly phase contrast, even simple previews of the data require a phase contrast method.

Live reconstruction shortens the time until feedback is available to the operators and can help to use the scarce beam time efficiently by re-adjusting or aborting measurements before they complete. Measurement times at PtyNAMi range between minutes for test measurements and many hours for large acquisitions.

Previously, Strauch et al.\ showed live \ac{SSB} ptychography~\citep{Strauch2021}. However, this algorithm is not suitable for this particular application since the reconstruction resolution is limited to the scan grid, and it can only reconstruct successfully if the scan step size is smaller than the resolving power of the illuminating beam, determined by its convergence angle. The low acquisition speed and resolving power of X-ray microscopes compared to electron microscopes means reconstruction with higher resolution than the scan grid, which allows coarse sampling of the specimen, is highly favorable here. Live \ac{WDD}~\citep{BangunWDD} has similar limitations.

Near live ptychographic reconstruction has also been shown by Pelz et al.~\citep{pelz2021real} by speeding up the reconstruction after fast scans so that it appears nearly instantaneous. However, this is  not applicable for this use case since the acquisition speed and not the reconstruction is the limiting factor, making a partial reconstruction during an ongoing scan desirable.

Using the well-established \ac{ePIE} implementation at DESY is favorable since it is in practical use for offline processing and the users' know-how is transferable. Towards this goal, three steps were undertaken:

\begin{enumerate}
    \item Create an API for the established stand-alone \ac{ePIE} implementation that allows to decouple data source, ptychographic processing and display of results.
    \item Investigate the behavior of iterative phase retrieval on partial input data in a simulated setting to determine if live iterative ptychography is possible for typical datasets acquired at PtyNAMi.
    \item Implement live ptychographic reconstruction at PtyNAMi.
\end{enumerate}

\section{Materials and Methods}

A reconstruction with an iterative method encompasses the following aspects:

\begin{itemize}
    \item Determine general parameters such as object and illumination size in pixel units, relation between physical coordinates and pixel coordinates.
    \item Preprocessing in order to match the different coordinate systems of object, illumination, scan and detector so that the forward and backward propagation model used in the optimization routine matches the actual measurements quantitatively.
    \item Assign initial values to object and illumination.
    \item Preprocess input data.
    \item Iterative optimization of illumination and object.
\end{itemize}

The following subsections detail how these aspects are adapted for live reconstruction.

\subsection{ePIE Implementation}

For the reconstruction of the experimental data, a GPU accelerated implementation of the \ac{ePIE}~\citep{MAIDEN20091256} algorithm is used. The update strength for object and illumination, typically denoted by $\alpha$ and $\beta$, was 1 for all reconstructions shown here.
The software is written in CUDA C++ and developed by the X-ray Nanoscience and X-ray Optics group at DESY for PtyNAMi.
The \ac{ePIE} is an iterative algorithm, which reconstructs the object and the illumination at the same time. 
The implementation starts with a general initialization phase which is then followed by the main reconstruction routine. 
The initialization phase prepares the diffraction patterns for the reconstruction and sets up the object and illumination with an initial guess. 
In this implementation the diffraction patterns are cropped to a size of $2^n$, with $n$ being an integral value, in order to use a simple fast-Fourier transformation and increase the performance.
Furthermore, the implementation can be configured to transform the diffraction patterns. Since the forward and back projection in this ePIE engine is implemented with a \ac{DFT}, the sampling and coordinate system of the reconstructed object transfer function must match the sampling and coordinate system of the diffraction patterns used in the reconstruction to allow forward and backward projection with a \ac{DFT}. In the ePIE implementation used at PtyNAMi, the spatial resolution and size of the object reconstruction in real space is determined by the size and scale of the diffraction patterns and the scan area. That means the spatial resolution and size of the object reconstruction can be adjusted by cropping the diffraction patterns before reconstruction. Furthermore, depending on the detector and its orientation, the recorded raw diffraction patterns might be rotated and/or flipped relative to the coordinate system of the position encoders. This can be compensated by rotating or flipping the diffraction patterns in such a way that their coordinate system and handedness match the position encoders.
In the second phase, the object and illumination are reconstructed. 
The implementation iterates over the diffraction patterns in a randomly shuffled order and calculates an update for the object and illumination.
The reconstruction ends after a specific number of iterations. 

\subsection{Extensions of the ePIE implementation}

The original \ac{ePIE} application is designed to perform an offline reconstruction, which means all diffraction patterns are available at the beginning of the reconstruction and the reconstruction configuration remains constant during runtime. 
There is no user input to add new diffraction patterns, and changes of the reconstruction parameters during the reconstruction are not possible. 
However, a live reconstruction requires that diffraction patterns can be added to follow the progressing scan during the reconstruction. 
Furthermore, it is necessary to stop and continue the reconstruction in order to extract intermediate updates, include new data and match the iteration speed with the acquisition speed to not over-optimize prematurely on a small subset of the input data.

For the interactive live reconstruction, Python bindings have been added to the existing C++ reconstruction engine using Pybind11 in order to connect it with the Python-based data handling and control logic. 
Compared to C++, Python provides a \ac{REPL} interpreter, which is perfectly suited for interactive workflows. 
Furthermore, Jupyter Notebooks can be used, giving some extra benefits such as an easy programming interface in the web browser and the possibility to combine source code with different media types, like markdown text and images to better document the reconstruction.

\subsection{Preprocessing}

Real-world detector data may require preprocessing such as cropping, scaling and orientation changes. Since the bindings don't support exporting or extending the loaded data, but require creating a new instance of the engine whenever the input data changes, a LiberTEM \ac{UDF} was implemented to perform preprocessing equivalent to the ePIE implementation based on its configuration file. This allows reusing the preprocessed data and reduces the amount of data loaded into the ePIE implementation. Furthermore, LiberTEM can perform this efficiently on live data streams in a parallel fashion.

\subsection{Determine parameters}

For offline processing of an entire dataset, the ePIE engine determines the reconstruction parameters from the size of the detector data in pixels after preprocessing, bounding box of physical scan positions, and optical parameters such as camera length, detector pixel size and wavelength from the configuration file. For live processing, the ePIE engine is initialized with mock detector data and approximate scan positions based on the scan settings that are known beforehand. The engine uses the nominal scan positions to calculate the nominal object size and pads it with half the illumination size so that the full area of the object that intersects with the illumination is covered. During the experiment, the scan positions are measured by interferometers or encoders.

Based on this initialization, a configuration for the engine is derived from the initial configuration that fixes the size and relation between physical coordinates and pixel coordinates. That way the input data can change dynamically during live processing without influencing the reconstruction parameters.

\subsection{Simulating live and parallel processing}

\begin{figure}
    \centering
    \includegraphics[width=180pt]{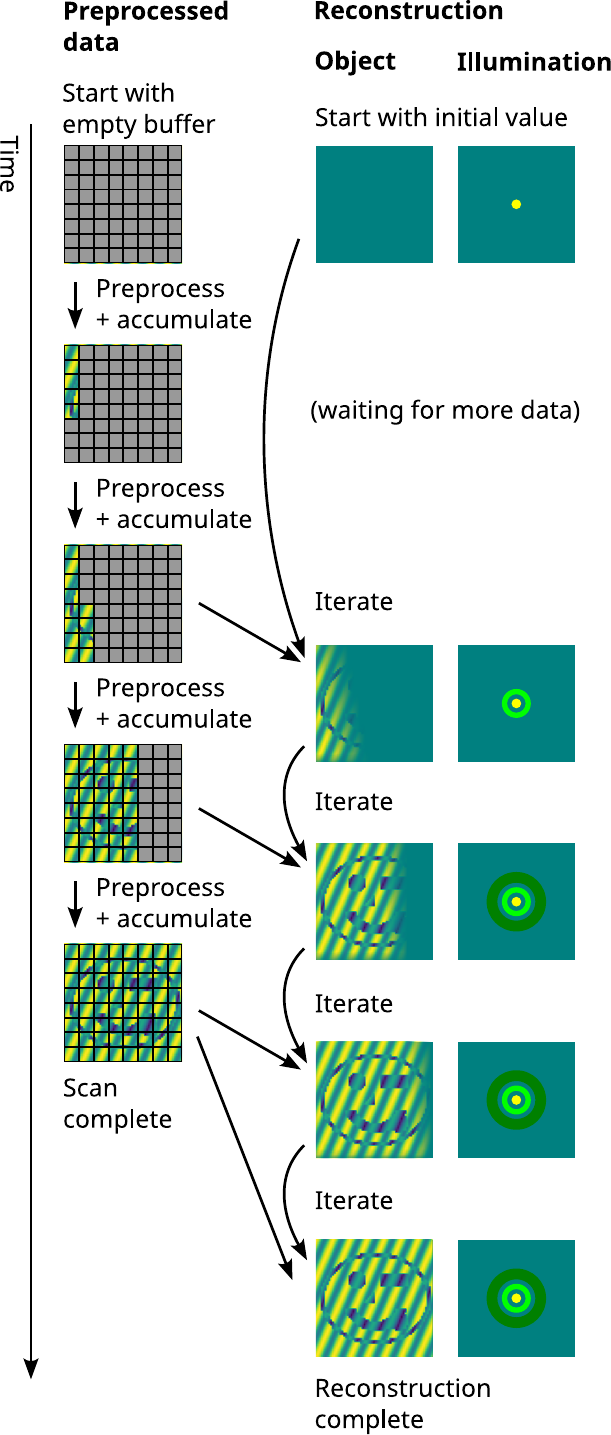}
    \caption{Schematic progress of live iterative ptychography. Data is preprocessed and accumulated, and the reconstruction algorithm includes more and more data in the iteration as it becomes available. This creates an evolving view of the object and illumination reconstruction. In real-world live processing, the sequence is determined by the acquisition speed, iteration speed and possibly limits on the number of iterations to avoid over-optimization. For simulated live processing, the sequence of available data and iteration steps is pre-determined by a script.}
    \label{fig:sequence}
\end{figure}

Live processing starts with a small initial subset of data that is extended gradually as the scan proceeds (Figure~\ref{fig:sequence}). As a preliminary test before implementing true live processing, this was simulated on offline data using the bindings described above: The initial values for object and illumination together with a small subset of the input data are loaded, a small number of iterations are performed, the resulting object and illumination are recorded, and the cycle is repeated with a larger subset of the input data using the previous state of object and illumination as initialization values until the complete dataset is consumed. Different scan schemes such as random sampling or spiral scans can be simulated by re-ordering the input data and positions.

Parallel processing can be simulated in a similar fashion by independent iteration on disjoint subsets of the input data with intermittent averaging between the object and illumination for synchronization, similar to~\citep{Nashed2014}. This creates a hybrid between classical ePIE where the object and illumination are updated dynamically with each diffraction pattern during a pass, and batch-oriented algorithms that accumulate updates and apply them only after a pass.

A ptychographic dataset for offline reconstruction had been recorded at PtyNAMi~\citep{Schropp:dy5005} on an NTT-AT resolution test chart with 50nm thinnest lines and spaces (model ATN/XRESO-50HC) at an X-ray energy of E = 13.5\,keV. The X-ray beam was focused by a pair of nanofocusing refractive lenses (NFLs)~\citep{Schroer_2005,Schropp_APL_2010} to a spot size of 116\,nm (horizontal) x 118\,nm (vertical). The sample was positioned slightly in defocus at a distance of approximately 200um. The diffraction patterns were collected with an Eiger X 4M detector (Dectris AG) positioned at a distance of 3.2m behind the sample. In the reconstructions these patterns were cropped to a size of (256, 256) pixels, which yielded in this case an effective real space pixel size of 15.3\,nm. The dataset was acquired in step-scan mode with a nominal scan window of 1\,$\mu$m x 1\,$\mu$m with a step size of 50\,nm. This gives 21 scan positions in vertical and horizontal scan direction, so that the resulting dataset consisted of 441 diffraction patterns in total, each with an exposure time of 0.5\,s. The approximate scan positions were obtained from the encoders of the piezo scanner and were used as a direct input for ptychographic phase retrieval. The position errors of the setup were not considered here. In practice, the beam, specimen and encoders can shift relative to each other.

The following processing schemes were implemented using this test dataset:

\begin{itemize}
    \item 1000 iterations on the full dataset for reference (Figure~\ref{fig:full}).
    \item Following the original line scan pattern in y direction, add 10 patterns and perform 10 iterations per step. After adding the complete data, iterate on the full dataset until 1000 iterations in total are reached (Figure~\ref{fig:gradual}).
    \item Add a random pick of 10 patterns and perform 10 iterations per step. Iterate on the full dataset until 1000 iterations in total are reached (Figure~\ref{fig:random}).
    \item Add 10 patterns following an inwards square spirangle pattern and perform 10 iterations per step. Iterate on the full dataset until 1000 iterations in total are reached (Figure~\ref{fig:spirangle}).
    \item Split the dataset into four random disjoint subsets. Perform 10 iterations on each subset independently, synchronize and repeat until 1000 iterations are reached. This simulates parallel processing on offline data (Figure~\ref{fig:parallel}).
    \item Add 10 patterns following the original line scan in y direction to four disjoint subsets in a round-robin~\citep{Kleinrock1964} fashion, perform 10 iterations on each subset after adding patterns to a subset, and synchronize. Iterate until 1000 total iterations are reached. This simulates parallel processing on live data (Figure~\ref{fig:parallelgradual}).
    \item Following the original line scan pattern in y direction, add 10 patterns and perform 1000 iterations per step to investigate impact of early over-optimization (Figure~\ref{fig:clipped}).
\end{itemize}

See section ``Data and code availability'' for the test data, ePIE implementation and test code.

\subsection{Live ePIE}

In the simulated live processing described above, the processing follows a predefined sequence of data loading, iteration and visualization that is executed sequentially in a deterministic fashion. All necessary parameters are read from configuration files.

\begin{figure}
    \centering
    \includegraphics[width=230pt]{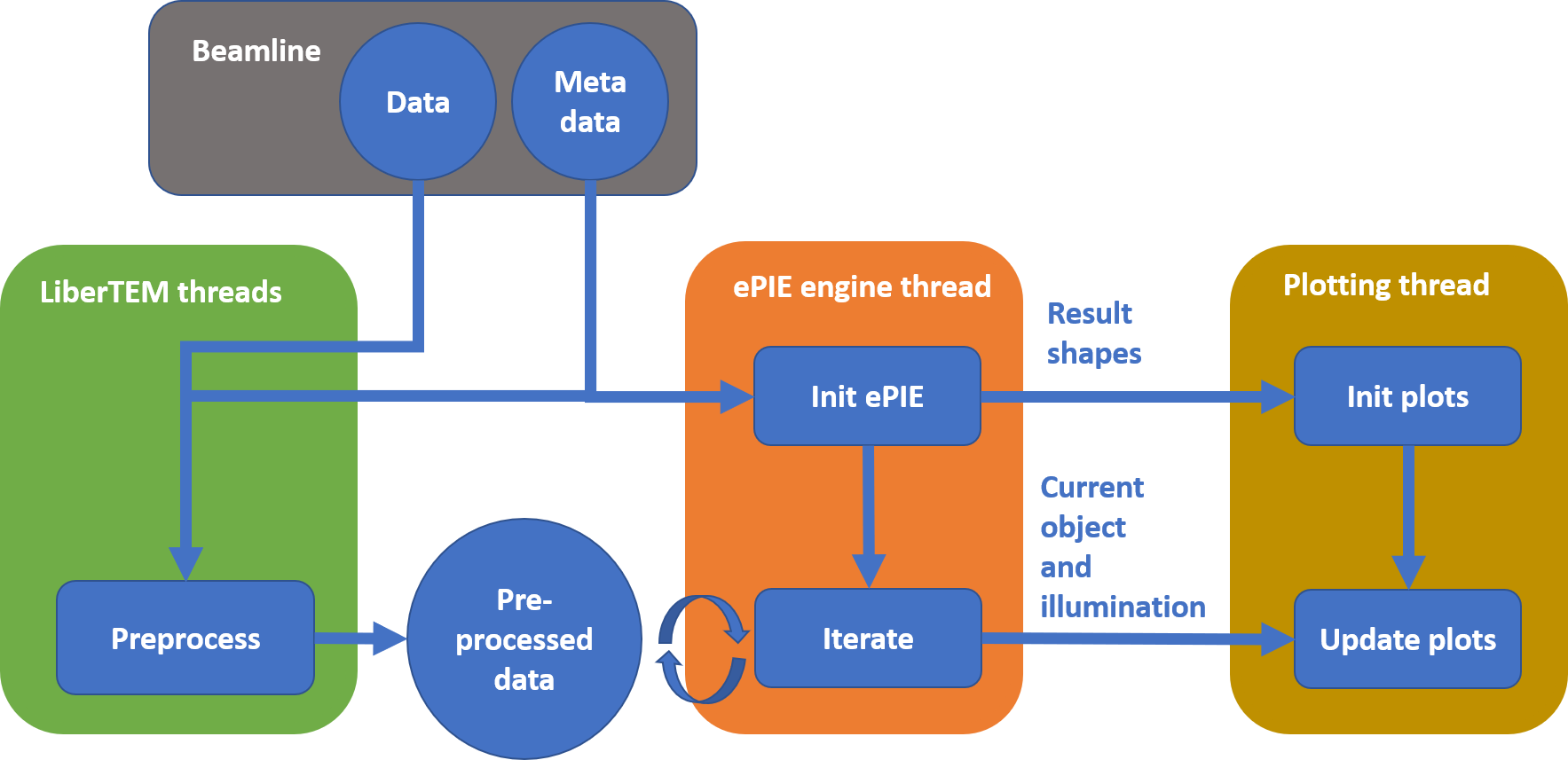}
    \caption{Diagram of the data flow for live ePIE.}
    \label{fig:diagram}
\end{figure}

After feasibility of live processing was demonstrated, these processes were implemented for real-world live processing at the DESY beamline in a Jupyter notebook as concurrent threads that follow the speed of the data source and iterate as fast as possible (Figure~\ref{fig:diagram}). It was executed on a node of the Maxwell cluster at DESY which has access to a cluster file system where results from the beamline are written in chunks with a short time delay, as well as the beamline's event interface. In principle, the data could also be streamed over network instead of using chunks in a file system, but such an interface is currently not implemented at the beamline. The pipeline is armed prior to data acquisition and starts processing as soon as a scan\_started event is received. Alternatively, the processing can attach to a running scan or load data from previous scans. The expected scan area is extracted from the scan\_started event, while other reconstruction parameters are determined from a configuration file for the ePIE engine.

A custom LiberTEM acquisition object was developed that reads the detector data from the folder structure that the beamline creates. This acquisition object is used as a data source for the preprocessing \ac{UDF}. In addition, another \ac{UDF} reads the matching stage encoder values from the beamline data to determine the position of each scan point. LiberTEM runs these \ac{UDF}s in a pool of worker processes in the background. The raw beamline data is available in~\citep{Weber2023}.

The run\_udf\_iter interface of LiberTEM provides a Python generator with updated partial preprocessing result from the \ac{UDF} for each received input data partition, together with a map that describes which parts of the expected data have been received so far. A feeder thread takes these updates from the generator and feeds them into the processing thread.

The processing thread consumes these updates and waits until a sufficient subset of data is received. Then it loads the data into the \ac{ePIE} implementation and starts iterating. Between iterations it checks if new data is available and loads it as required, and regularly feeds an intermediate state of the reconstruction into a queue for display.

The visualization thread reads from this display queue and updates LiberTEM live plots with each new result.

Live processing was performed while recording data on an NTT-AT resolution test chart with 50nm thinnest lines and spaces (model ATN/XRESO-50HC) at an X-ray energy of E = 9\,keV. The X-ray beam was focused by a Fresnel zone plate (FZP) with outer diameter of 125\,$\mu$m and thinnest outermost lines of 70\,nm. At the employed X-ray energy, these optics create a focus with a size of about 85\,nm at a focal distance of 64\,mm. The diffraction patterns were collected with an Eiger X 4M detector (Dectris AG) positioned at a distance of 3.265\,m behind the sample. In the reconstructions, these patterns were cropped to a size of (256, 256) pixels. The dataset was acquired using a continuous meander scan with a nominal scan window of 10\,$\mu$m x 11\,$\mu$m and a step size of 100\,nm. The resulting dataset consisted of 11245 diffraction patterns in total, each with an exposure time of 20\,ms.

The test object was mounted in a different orientation between the measurement campaign for the simulated live processing and the actual live processing. Both the recording for simulated live processing and for real live processing were not adapted specifically for this purpose, but performed alongside regular beamtimes at PtyNAMi during the preparation phase for real measurements. The parameters are different and in the typical range for PtyNAMi for that reason.

\section{Results}

\subsection{Simulated live and parallel processing}

\begin{figure}
    \centering
    \includegraphics[width=230pt]{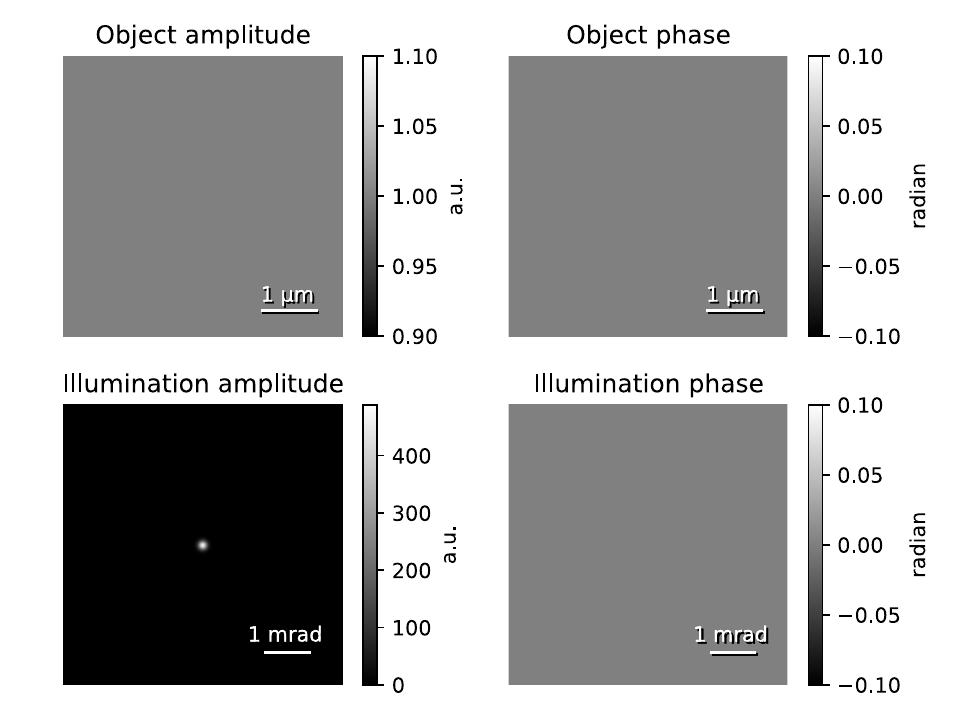}
    \caption{Starting value for all simulated live reconstructions. The object is initialized with the identity function. The illumination amplitude is initialized with a small Gaussian peak and flat phase of 0.}
    \label{fig:starting}
\end{figure}

\begin{figure}
    \centering
    \includegraphics[width=230pt]{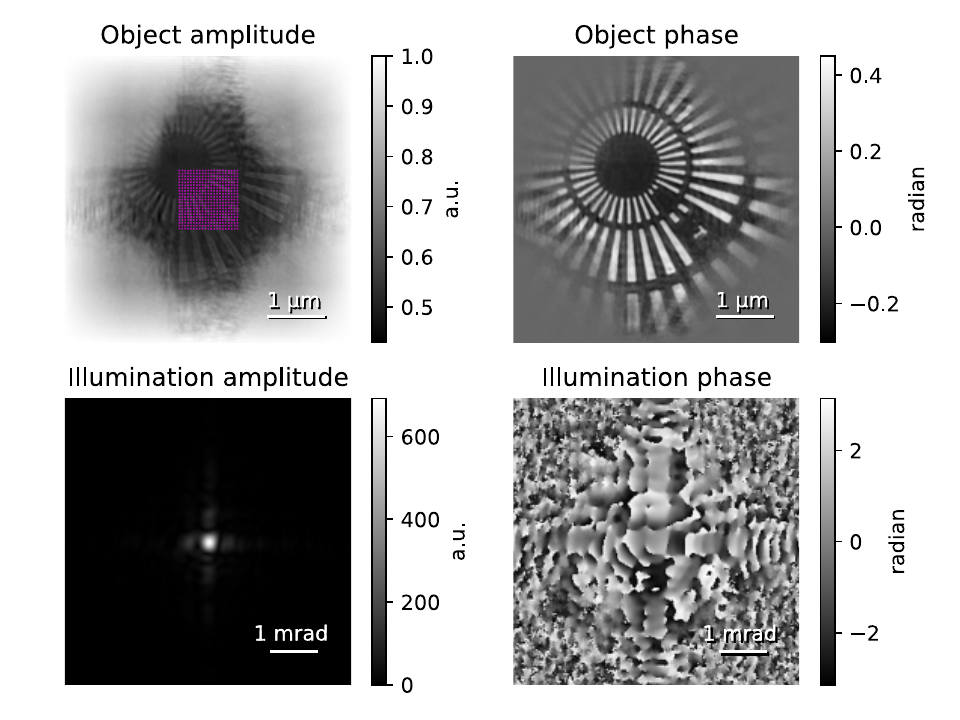}
    \caption{Reference result with 1000 iterations on the full dataset. The magenta dots in the upper left plot indicate the scan positions.}
    \label{fig:full}
\end{figure}

Figure~\ref{fig:starting} shows the starting value of illumination and object used in all simulated live processing.

Figure~\ref{fig:full} shows amplitude and phase of the reconstruction of both illumination and object using 1000 iterations on the complete dataset, corresponding to conventional offline processing.

\begin{figure}
    \centering
    \includegraphics[width=230pt]{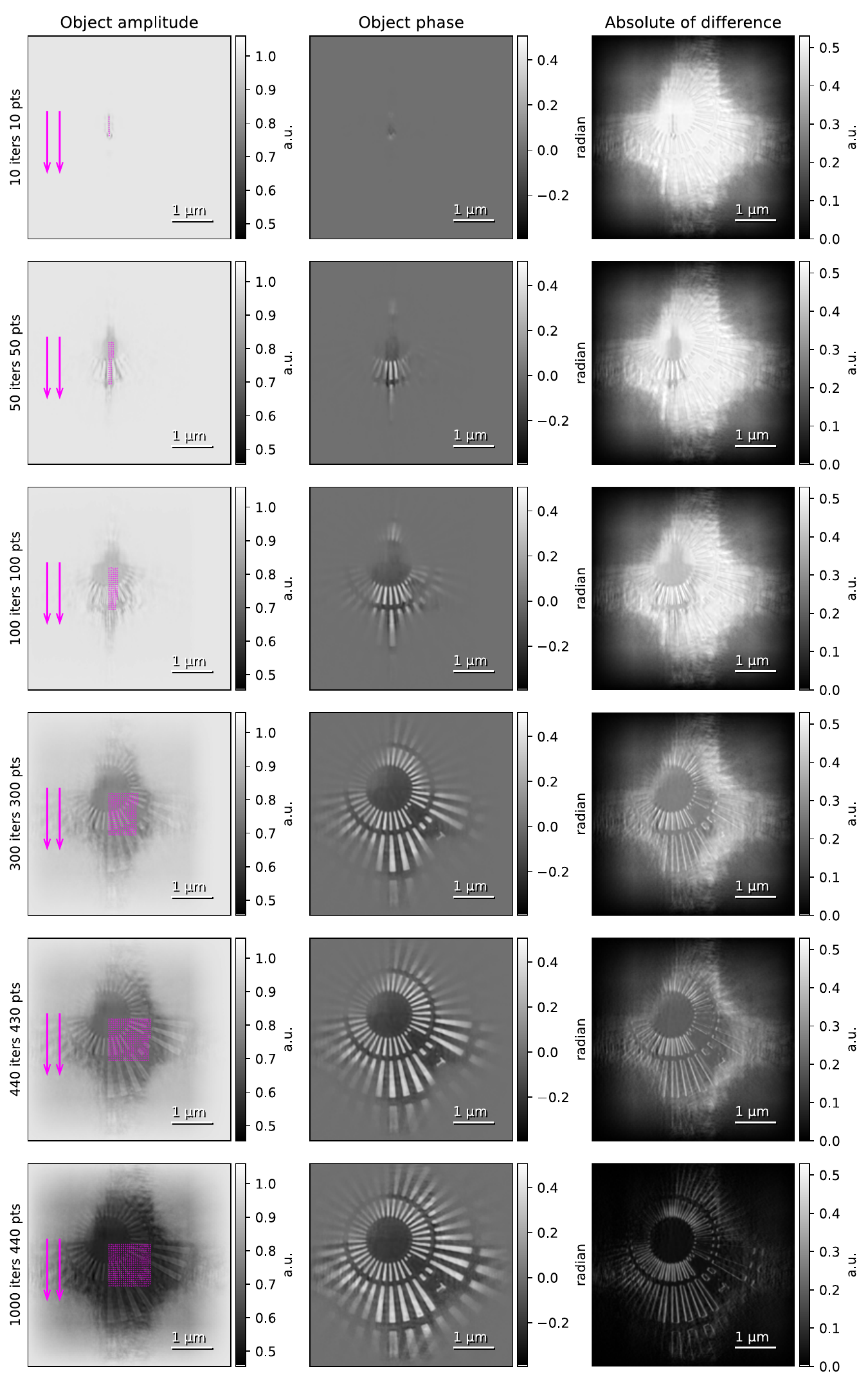}
    \caption{Simulated live reconstruction following the original line scan pattern in y direction. The magenta dots in the left column indicate the scan positions. The right column shows the absolute value of the difference to the reference result.}
    \label{fig:gradual}
\end{figure}

\begin{figure}
    \centering
    \includegraphics[width=230pt]{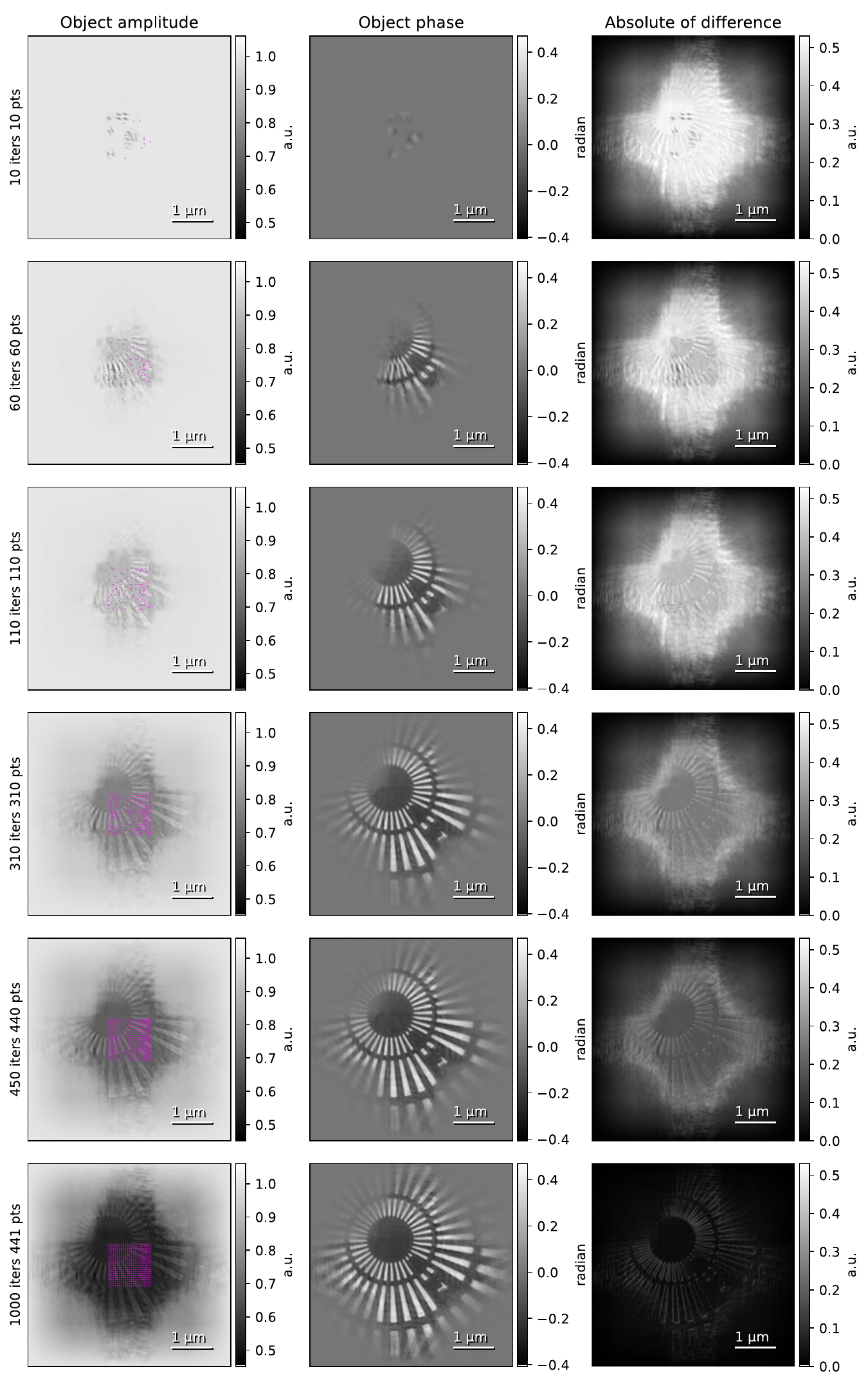}
    \caption{Simulated live reconstruction with a random scan pattern. The magenta dots in the left column indicate the scan positions. The right column shows the absolute value of the difference to the reference result.}
    \label{fig:random}
\end{figure}

\begin{figure}
    \centering
    \includegraphics[width=230pt]{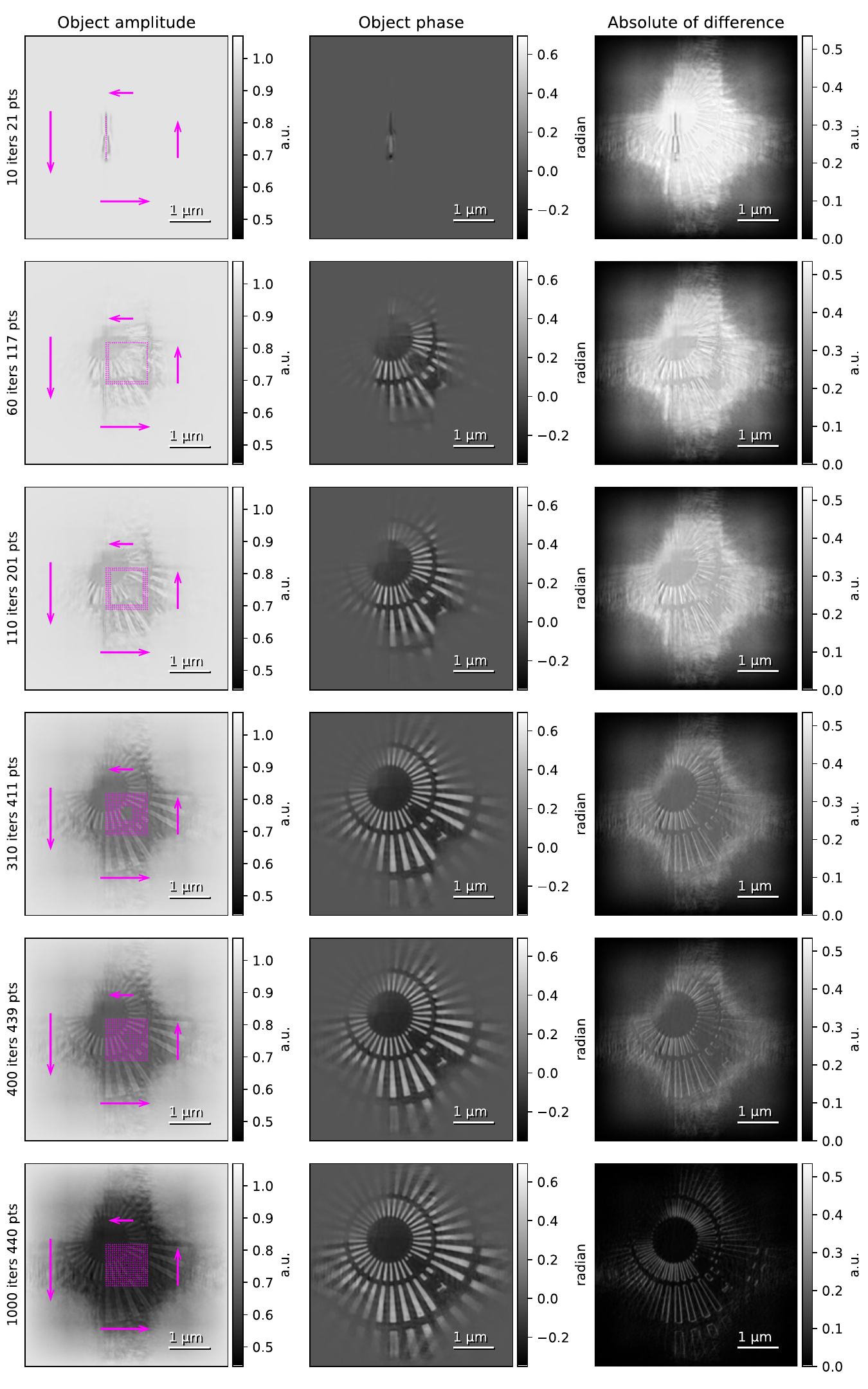}
    \caption{Simulated live reconstruction following a spirangle scan pattern. The magenta dots in the left column indicate the scan positions. The right column shows the absolute value of the difference to the reference result.}
    \label{fig:spirangle}
\end{figure}

\begin{figure}
    \centering
    \includegraphics[width=230pt]{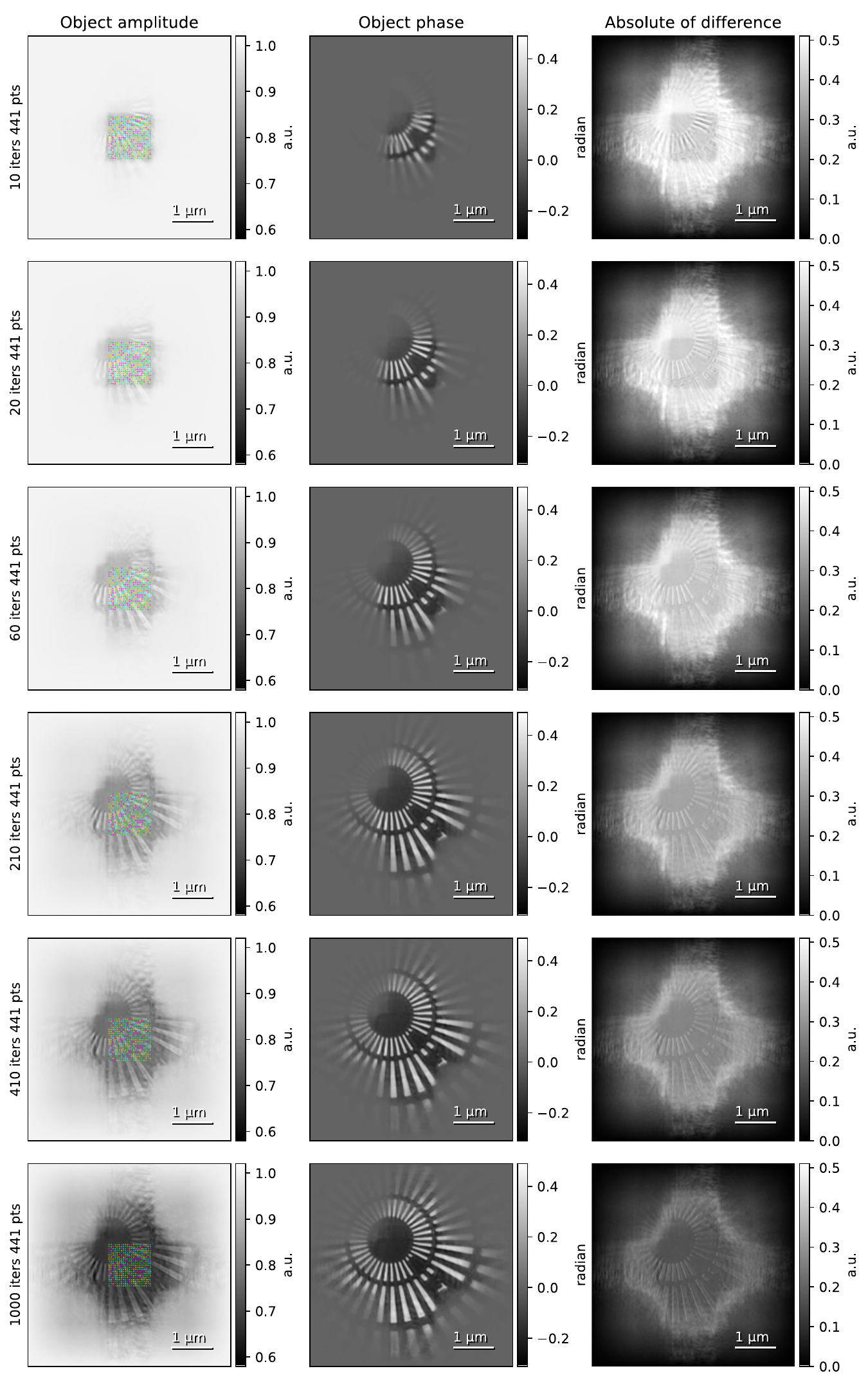}
    \caption{Simulated parallelized reconstruction on four random disjoint subsets. The colored dots in the left column indicate the scan positions, with the color indicating the subset. The right column shows the absolute value of the difference to the reference result.}
    \label{fig:parallel}
\end{figure}

\begin{figure}
    \centering
    \includegraphics[width=230pt]{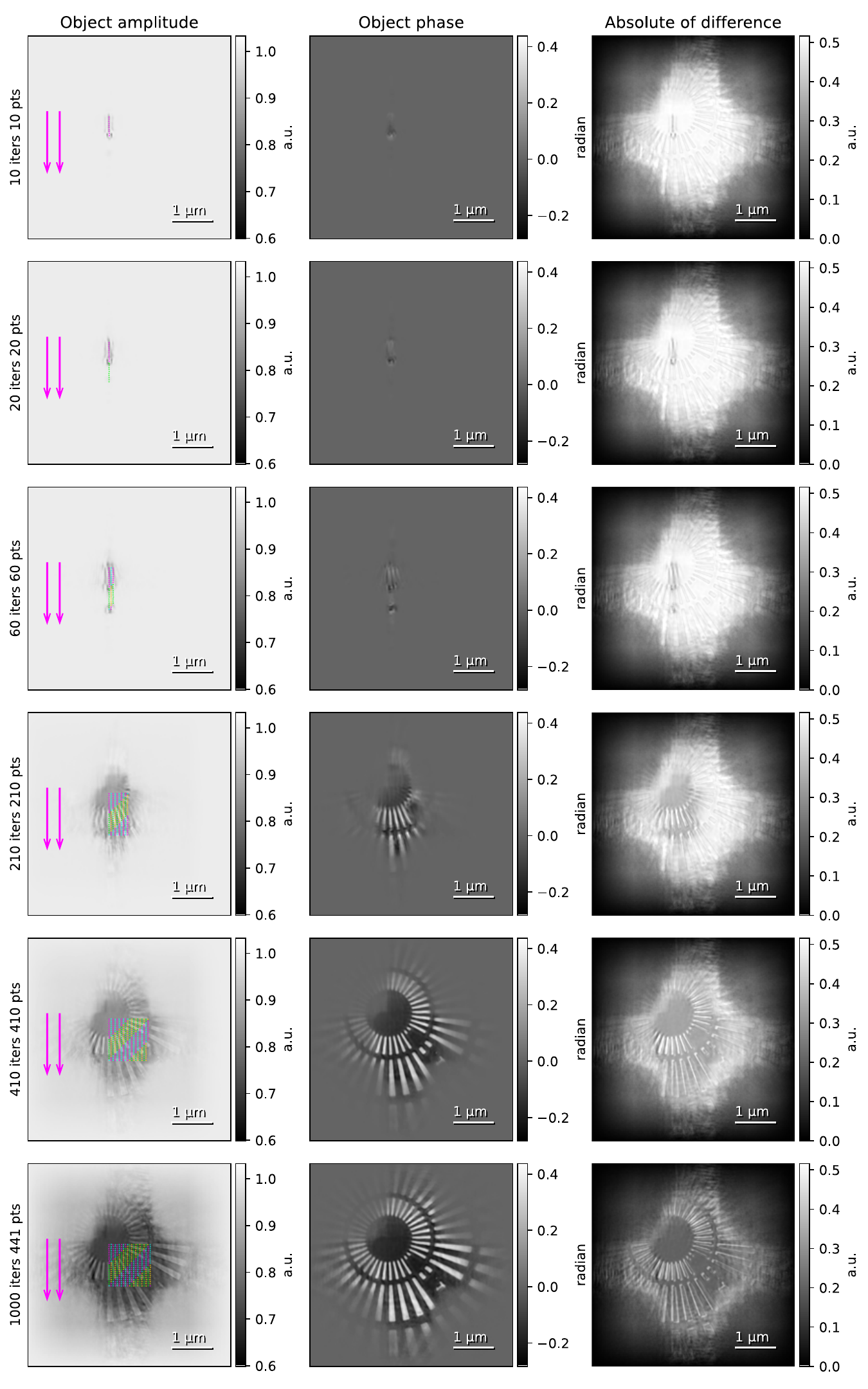}
    \caption{Simulated parallelized live reconstruction on four gradually growing disjoint subsets. The colored dots in the left column indicate the scan positions, with the color indicating the subset. The right column shows the absolute value of the difference to the reference result.}
    \label{fig:parallelgradual}
\end{figure}

Figures~\ref{fig:gradual} to~\ref{fig:parallelgradual} show how the simulated live and parallel processing results evolve with different processing and subdivision schemes. Note how the reconstruction extends far beyond the scan range due to the extended size of the illumination.

In all cases a reconstruction with strong resemblance to the final result emerges already after a few iterations on a small subset of the dataset. The random (Figure~\ref{fig:random}) and spirangle (Figure~\ref{fig:spirangle}) scan patterns cover the entire scan area early on, demonstrating how such acquisition schemes can be advantageous for live processing compared to a sequential line scan to get an overview of the object early on.

\begin{figure}
    \centering
    \includegraphics[width=230pt]{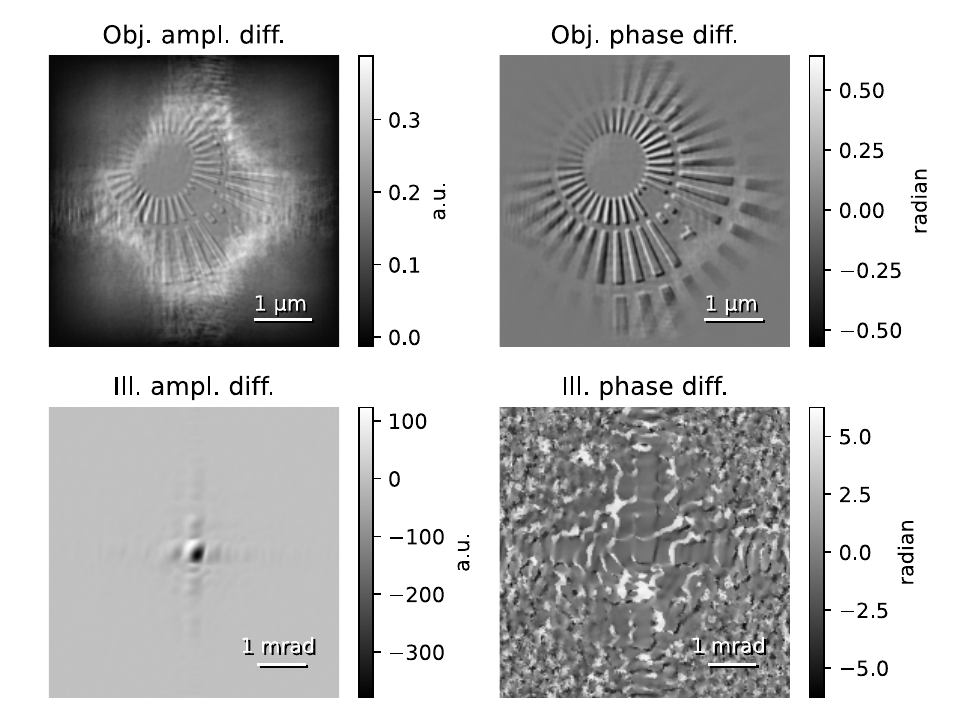}
    \caption{Amplitude and phase difference between the final result of Figure~\ref{fig:parallelgradual} and the reference result (Figure~\ref{fig:full}). The amplitude of the object is higher, while the amplitude of the illumination is lower. The contour line around the sharp features of the object phase indicate a small shift relative to the reference result.}
    \label{fig:parallel_gradual_diff}
\end{figure}

After completing the same number of iterations, the reconstructions that are not parallelized differ from the reference result mainly by a small shift that manifests itself as a contour line around sharp features of the object in the difference plot. The parallelized results exhibit a stronger difference that did not vanish up to 1000 iterations. The plot of amplitude and phase difference in Figure~\ref{fig:parallel_gradual_diff} reveals that the amplitude of the object is larger and the amplitude of the illumination is lower than the reference. However, they still show a qualitative visual resemblance to the reference result.

\begin{figure}
    \centering
    \includegraphics[width=230pt]{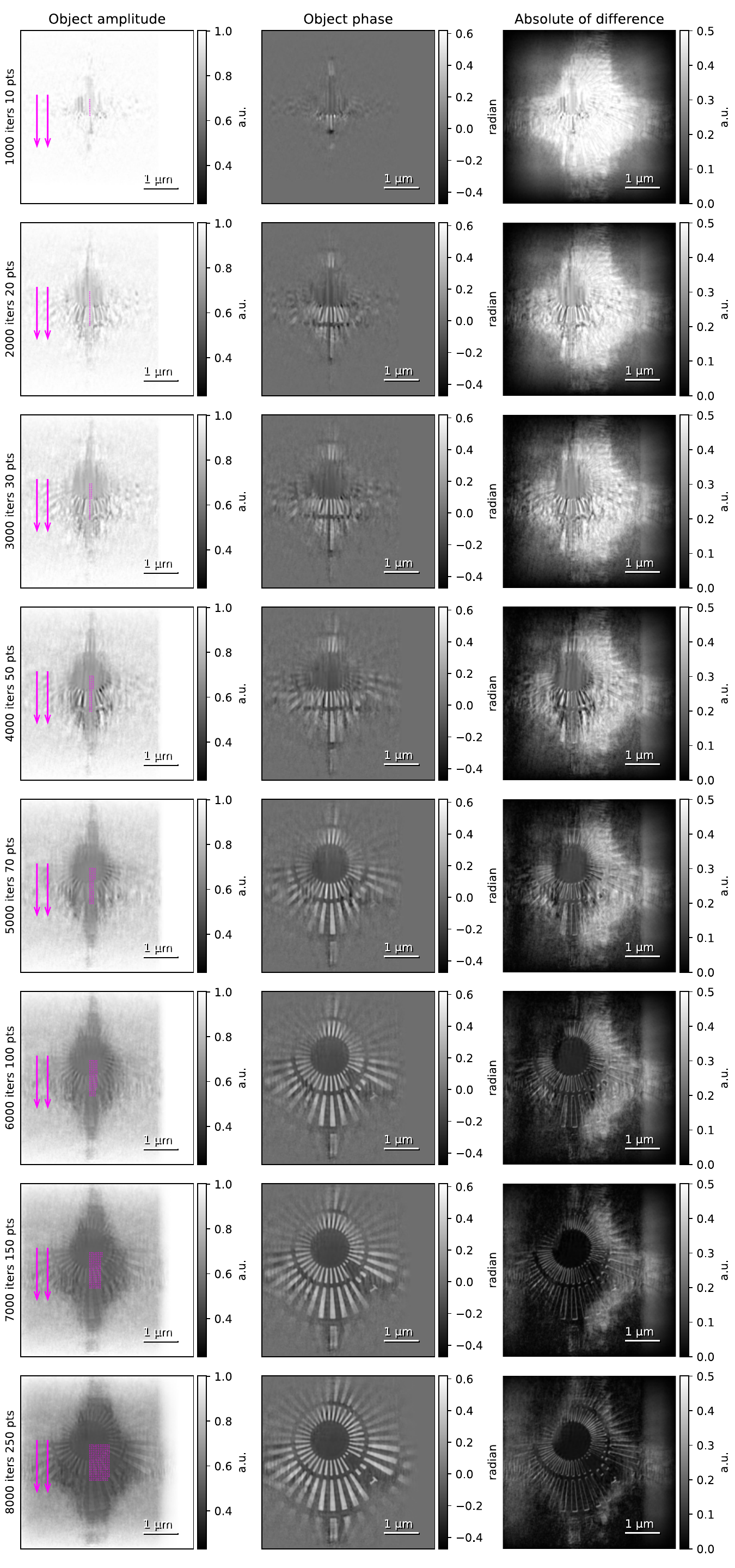}
    \caption{Simulated live reconstruction following the original line scan pattern in y direction with over-optimization on a small subset of input data. The magenta dots in the left column indicate the scan positions. The right column shows the absolute value of the difference to the reference result.}
    \label{fig:clipped}
\end{figure}

Animated videos of the simulated live reconstruction are available in~\citep{Weber2023}.

\subsection{Live processing}

\begin{figure}
    \centering
    \includegraphics[width=230pt]{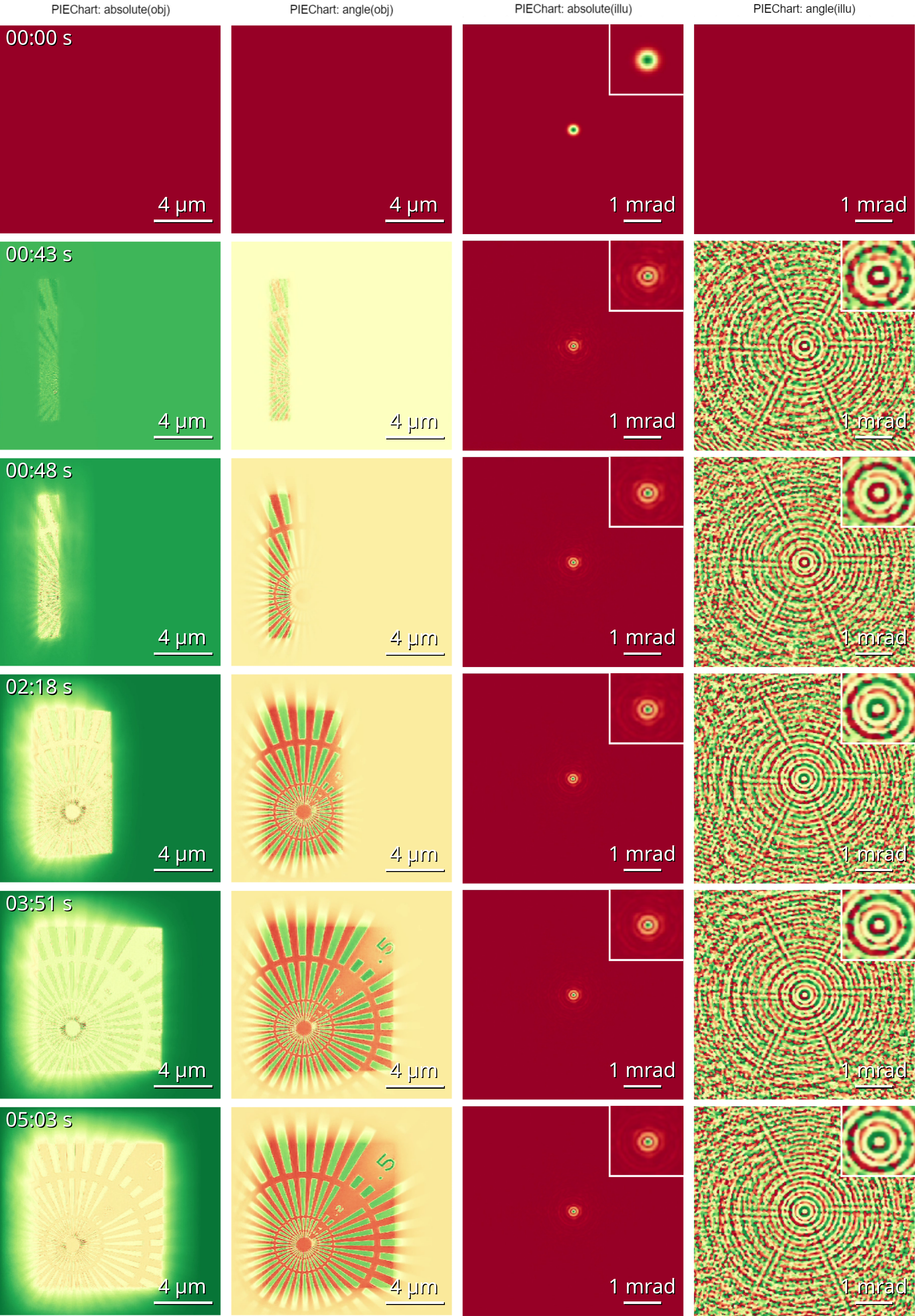}
    \caption{Evolution of amplitude and phase for object and illumination that was recorded during live processing. The screen with the Jupyter notebook running the reconstruction was recorded as a video. This Figure shows rearranged and annotated snapshots of the plots taken at the indicated times relative to the start of the scan. The full video is available in the Supplementary Material. The insets show the center of the illumination scaled by a factor of 2. The first step shows the initialization before iteration starts. The second step was recorded just after the start of iteration. Steps 3 and 4 show arbitrary intermediate states. The second to last step shows the state at the moment the scan completes. The last snapshot was taken after a number of additional iterations. In the beginning, 50 iterations were executed for each batch of 500 images received from the beamline. Towards the end, about 6 iterations were executed per batch due to the much larger number of scan points included in each iteration. The live plots were updated every two iterations for quick visual feedback.}
    \label{fig:reallive}
\end{figure}

Figure~\ref{fig:reallive} shows how the live view of object and illumination reconstruction develops during live processing. This allows monitoring the scan. Since the reconstruction can be restarted and re-attached to a running scan, users can already start optimizing reconstruction parameters while a scan is running. See ``Data and code availability'' regarding availability of source code, raw data and animated results.

A screen capture video of the live reconstruction is available in the supplementary material.

\section{Discussion}

The results on simulated and real-world live processing show that iterative ptychography algorithms can be used for live reconstruction by starting the iteration on a small subset of the data that grows as the data acquisition proceeds. The state of object and illumination are carried forward. For this application a ptychography engine should offer the following features:

\begin{itemize}
    \item An API that allows integration in a complex workflow, as opposed to stand-alone programs.
    \item Flexible and fast import and export of object, illumination and data via the API.
    \item Stepping the iteration to extract intermediate values or stop as required.
    \item Options to encourage stable convergence, such as good starting values for illumination and object, restrictions on value ranges and illumination shapes, countering impact of noise and/or detector gain characteristics, masking or cropping invalid pixels, etc.
    \item Configuration of a stable, pre-determined size and shape of the object and illumination as well as a relation between physical scan position and pixel position in the reconstructed object based on the expected acquisition parameters.
    \item Separation of preprocessing from the iteration engine.
\end{itemize}

In particular when the beamline at DESY acquires data at a low rate, around 10 frames per second, it is very advantageous to obtain a view of the object early on during an acquisition to abort and adjust parameters in case issues are detected, such as the choice of acquisition area, exposure, or focal plane. During real-world live processing the implementation could easily keep up with this data rate. During the experiment shown in Figure~\ref{fig:reallive}, the region on the test specimen was already clearly recognizable after completion of about a quarter of the scan. An operator has a chance to detect issues with the field of view or settings at this point. Since large scans at PtyNAMi can run for many hours overnight, detecting issues at 25\% completion or earlier can save significant amounts of beam time and allows early optimization of reconstruction parameters.

Figure~\ref{fig:clipped} shows that the reconstruction doesn't diverge even with a high number of iterations on a small initial subset for this particular dataset, engine and choice of parameters. As soon as a region of the test specimen is covered by a sufficient number of scan points, the iteration stabilizes on a value similar to the reference result in this region. Similar to the other scans, this reconstruction exhibits a small shift that results in a bright line around sharp features in the difference plot. A result that clearly shows features of the test object was obtained with three and a half scan rows in this test. This indicates that a successful reconstruction requires some degree of lateral coverage with overlapping illuminations, but starting the iteration early with a smaller number of diffraction patterns is not harmful in this case. The lower limit for successful partial reconstruction will likely be different for different illumination shapes and sizes, scan patterns, objects, and reconstruction engine settings. Typically, engines for iterative ptychographic reconstruction will have many parameters and constraints for regularization in order to ensure a stable convergence with real-world data. Here, we merely show that the results for live processing for all tests that we performed eventually converged towards the offline result as more and more data was added.

As long as the convergence is stable, meaning it converges quickly towards the same result independent of the starting values for illumination and object, and stabilizes after a sufficient number of iterations, live reconstruction will eventually converge on the offline result under all circumstances: The previous reconstruction steps with a subset of the input data merely generate a starting value for illumination and object for the final iteration phase on the complete input data. The final iteration phase will then converge and stabilize on the offline result.

Furthermore, any real-world dataset for ptychographic reconstruction can be understood as a partial dataset: The actual scan size and the size of the object are only limited by practical concerns and can be, in principle, arbitrarily large. That means any real-world dataset can be extended by just continuing the scan beyond the initial borders, making the initial ``full'' dataset a partial dataset. The reconstruction on a subset of a large dataset is not distinguishable from the reconstruction of a ``full'' dataset with the same scan coordinates as the subset. Regions of the object transfer function that are not touched by the illumination function for any scan position just stay at their initial values.

The results shown here conform to this mathematical expectation: Eventually, a reconstruction on a growing subset of a sufficiently large dataset will give a stable reconstruction for all regions of the object that are covered by the illumination function in that subset, provided the reconstruction of the complete dataset is stable. For a general discussion and further references on the convergence of ptychographic reconstruction see, for example,~\citep{Melnyk2023}.

Simulated live reconstruction and actual live reconstruction mainly differ in their input and output method: In simulated live processing, a ``recipe'' with well-defined steps is executed, and all intermediate results are stored and then plotted. In actual live processing, the sequence of data and iteration is determined by the interplay of data source, preprocessing and iteration thread. The intermediate results are not recorded here, but just plotted for quick visual feedback. Notably, here live reconstruction is not meant to replace careful offline reconstruction with optimized parameters. It is rather a quick preview tool to observe what data is being recorded and to check if the chosen parameters are likely to give a successful offline reconstruction later.

In this paper, we show practical evidence that such live reconstruction can be possible and beneficial under the typical conditions at PtyNAMi. In particular, live reconstruction worked reliably for different illumination shapes and scan parameters that were chosen somewhat arbitrarily from typical values. The live reconstruction seems rather robust against variations in the set of diffraction patterns and number of iterations: By adding more and more diffraction patterns, the result reliably converged to closely resemble the offline reconstruction result in the region covered by the scan, meaning it seems to meet the conditions for successful partial reconstruction described in the previous paragraphs. It seems intuitive that a scan that yields a satisfactory reconstruction with only a subset of scan points is likely to also yield a satisfactory result for the full scan. That means it is valuable as a preview tool to give an early indication to operators if the scan and reconstruction parameters will give a good final result. Beamline operator feedback was positive.

The parallelized reconstruction schemes can potentially be used to generate faster previews by using multiple GPUs in parallel. An engine and processing scheme that produces the same result independent of the number of parallel GPUs is required, which was not yet achieved here. Generally, algorithms that act in a parallel manner, such as \ac{RAAR} or \ac{DM}, are advantageous for this~\citep{Enders2016}. The PtyPy software package~\citep{Enders2016} already offers parallelized reconstruction and can potentially be used as a reconstruction engine for live processing in a similar fashion.

\section{Outlook}

Faster processing seems feasible, in particular targeting electron ptychography with fast direct detection detectors and fourth generation synchrotron radiation sources such as PETRA IV, since the preprocessing with LiberTEM is efficient and highly scalable, and parallelized live processing on several GPUs seems feasible with a suitable reconstruction engine. Further optimization of the ePIE engine is thinkable as well, for example to run on different platforms such as field-programmable gate arrays (FPGAs) using Alpaka~\citep{10.1007/978-3-319-67630-2_36}.

In the future, live reconstruction can be factored in when choosing scan patterns. A spiral or spirangle seems promising since it combines short movements of the stage between subsequent acquisitions with early coverage of the whole field of view, and such patterns are already available at PtyNAMi. A conventional line scan pattern can be advantageous if the position encoders exhibit hysteresis. Further experiments to test the behavior of live reconstruction for different scan patterns, specimens, illuminations and scan step sizes seem promising. Here, we demonstrate that live reconstruction can yield useful results for different scan patterns.

The approach for live iterative ptychography described here is not limited to ePIE, but should work with any reconstruction engine that offers a similar interface. Implementations of different reconstruction approaches such as multislice ptychography, mixed state ptychography, near field ptychography~\citep{You2023} and/or alternative projection schemes such as \ac{RAAR}, \ac{ADMM} and similar can meet the requirements for live processing described in the previous section: They optimize gradually, and the set of scan positions is flexible. Furthermore, this approach should be transferable to other experimental modalities such as extreme ultraviolet light, visible light and electrons. A trial with electron microscopy data is included in the Supplementary Material.

Live monitoring is not limited to iterative ptychographic reconstruction: Any other processing or reconstruction method that is compatible with LiberTEM can be included with little effort in the preprocessing stage, such as Centre of Mass or even direct ptychographic reconstruction~\citep{Strauch2021,BangunWDD}.

Furthermore, in the future it can be investigated under which conditions the reconstruction is quantitative and how large the errors are at various stages of the scan and reconstruction.

Panning and zooming with a live view may be possible by applying a suitable transformation to an intermediate state of the object in order to generate a new starting value, generating a new configuration and continuing the acquisition and iteration with the new parameters. This can allow live navigation on the specimen and adjustment of parameters at the beam line, effectively creating an interactive microscope for X-rays and other ptychography modalities. The results also show that non-standard scan patterns can optimize early coverage of the field of view. Previously, Velazco et al.~\citep{VELAZCO2020113021} have shown non-standard scan strategies, including a space-filling curve. Live reconstruction can also enable adaptive scanning strategies~\citep{Ede2021}.

\begin{acronym}[Ptycho]
    \acro{ePIE}{extended ptychographical iterative engine}
    \acro{REPL}{read–eval–print loop}
    \acro{UDF}{user-defined function}
    \acro{SSB}{single side-band}
    \acro{WDD}{Wigner distribution deconvolution}
    \acro{ADMM}{alternating direction method of multipliers}
    \acro{RAAR}{relaxed averaged alternating reflections}
    \acro{DM}{difference map}
    \acro{DFT}{Discrete Fourier Transform}
\end{acronym}

\section{Data and code availability}

The ePIE reconstruction engine and interface for live processing is available upon reasonable request from Andreas Schropp \href{mailto:andreas.schropp@desy.de}{andreas.schropp@desy.de}.

Additional movies that show the live reconstruction, the dataset and notebook for simulated offline processing, the Jupyter notebook for live processing at the beamline, and the beamline data recorded during the live processing run are available at \url{http://doi.org/10.5281/ZENODO.8239052}~\citep{Weber2023}.

\section{Competing interests}

The authors declare no competing interests.

\section{Acknowledgements}

\begin{itemize}
    \item Parts of this research were carried out at beamline P06 of PETRA III at Deutsches Elektronen-Synchrotron DESY, a member of the Helmholtz Association (HGF).
    \item We acknowledge support from Helmholtz Association under contract No.~ZT-I-0025 (Ptychography 4.0) and  JL-MDMC (Joint Lab on Model and Data-Driven Material Characterization).
    \item This research was supported in part through the Maxwell computational resources operated at Deutsches Elektronen-Synchrotron DESY, Hamburg, Germany
    \item This work was partly funded by the Center for Advanced Systems Understanding (CASUS) which is financed by Germany's Federal Ministry of Education and Research (BMBF) and by the Saxon Ministry for Science, Culture, and Tourism (SMWK) with tax funds on the basis of the budget approved by the Saxon State Parliament.
\end{itemize}

\end{document}